\documentclass{article}

\begin{document}

\title{Permutation Symmetry For Many Particles}

\author{M. J. York\\975 S. Eliseo Dr. \#9, Greenbrae, CA 94904, USA\\{\it email}: mikeyork@home.com}      

\date{Dec 12, 1999}          
\maketitle
\begin{abstract}
We consider the implications of the {\em Revised Symmetrization Postulate\footnote{See ref \cite{York}}} for states of more than two particles. We show how to create permutation symmetric state vectors and how to derive alternative state vectors that may be asymmetric for any pair by creating asymmetric interdependencies in their state descriptions. Because we can choose any pair to create such an asymmetry, the usual generalized exclusion rules which result, apply simultaneously to any pair. However, we distinguish between simultaneous pairwise exclusion rules and the simultaneous pairwise anti-symmetry of the conventional symmetrization postulate. We show how to construct a variety of state vectors with multiple interdependencies in their state descriptions and various exchange asymmetries --- including one which is anti-symmetric under exchange of two bosons --- all without violating the spin-statistics theorem. We conjecture that it is possible to construct state vectors for arbitrary mixes of bosons and fermions that emulate the conventional symmetrization postulate in a limited way and give examples. We also prove that it is not possible to define a single state vector that simultaneously obeys the conventional symmetrization postulate in its standard form (in which the exchange phase does not depend on the spins of additional particles that are present) for every pair that can be interchanged.
\end{abstract}

\section{Introduction}      
In a recent paper\cite{Tino}, Tino has suggested that to provide an experimental test for the ``symmetrization postulate'', as opposed to simply test the spin-statistics theorem, it is necessary to look at states of three or more identical particles. 

However, in a previous paper\cite{York}, we have demonstrated the short-comings of the {\em conventional} symmetrization postulate (hereafter referred to as the CSP) which claims that states of identical fermions are necessarily anti-symmetric. Specifically we have proved that, given two axioms relating to (i) permutation not being a physically significant transformation and (ii) uniqueness of state vectors given a physically complete state description (which, amongst other things, involves eliminating the possibility of arbitrary $2\pi$ rotations of the spin quantization frame), then it is always possible, by describing individual states in an order independent way (including independent, but symmetrically chosen, spin quantization frames for each particle), to construct state vectors that are pure permutation symmetric regardless of whether the particles are identical or not and regardless of their spin. We called this the {\em Revised Symmetrization Postulate} (RSP).

We also showed that this finally makes possible a complete proof of the spin-statistics theorem on purely geometrical grounds, without recourse to either field theory (see \cite{Duck&Sud1}\cite{Duck&Sud2} for a summary), or relativistic S-matrix analyticity\cite{Stapp} or specific choices of exchange transformation\cite{Broyles}\cite{Bacry}\cite{Berry&Rob}. The conventional fermion anti-symmetry, in the two-particle case, is shown to result from an implicit asymmetry in the individual state descriptions which leads to a rotation by $2\pi$ on the spin quantization frame of one particle relative to the other when the asymmetry is reversed by the ``exchange'' operation. Since such anti-symmetric state vectors are related to the permutation symmetric state vectors, by a simple (but order dependent) phase factor the symmetric state vectors also vanish under exactly the same circumstances as the anti-symmetric state vectors. 

Since the CSP is now proven --- in the two-particle case, at least --- to be just a convention that contains no new physics, any of the tests suggested by Tino, should now be viewed as a test of the RSP (or our axioms of permutation invariance and uniqueness for physically complete state descriptions). For this reason we shall examine, in this paper, the consequences of the RSP for states of three or more particles.

With this in mind, we show how to construct state vectors for states of more than two particles, that are symmetric under pure permutation of any pair, whatever their spin, by employing independent (but symmetrically chosen) spin quantization frames. We then examine the consequences of transforming to a system in which all particles have their spin quantized in a common canonical frame and thereby show how to create pairwise order dependence in the state vector. This enables us to prove simultaneous pairwise exclusion rules for all pairs of particles with identical quantum numbers in the multi-particle state. 

However this does not mean that we can define state vectors that are simultaneously anti-symmetric for all pairs of identical particles simultaneously. This is a much more restrictive condition. To see this, consider the case of three identical particles and suppose that we can create three distinct state vectors that each obey a distinct anti-symmetry for a different specific pair:
\begin{eqnarray}
|S^1_i;S^2_j;S^3_k>^k & = & - |S^1_j;S^2_i;S^3_k>^k\nonumber\\
|S^2_i;S^3_j;S^1_k>^j & = & - |S^2_k;S^3_j;S^1_i>^j\nonumber\\
|S^3_i;S^1_j;S^2_k>^i & = & - |S^3_i;S^1_k;S^2_j>^i
\end{eqnarray}
where the superscript identifying the ket indicates which particle is {\em not} exchanged. If each of these state vectors describes the same state, they can differ by, at most, a phase factor. If one vanishes when two states become indistinguishable (e.g. $S_1=S_2$), because of the anti-symmetry under a particular exchange, then they must all vanish. However, in the general case, of distinguishable states, there need not be a single state vector (i.e. where we can ignore the superscript labeling the ket) that obeys all three anti-symmetries simultaneously. The CSP, on the other hand, claims that there {\em must be} a single state vector that is simultaneously anti-symmetric under all exchanges of identical fermion pairs, and, in its standard form, requires this {\em regardless of the presence of other particles or their spins}.

One of the chief purposes of this paper is to explore whether or not the RSP (i.e. pure permutation symmetry) can lead to the construction of a state vector that obeys the CSP. It will be our conclusions that:
\begin{enumerate}
\item It is possible to construct a multi-particle state vector that {\em emulates}, in a limited way, the property of simultaneous exchange antisymmetry under fermion exchange, but only by creating highly complex asymmetries in the individual state descriptions of the fermions present.
\item It is not possible, for states of three or more fermions, to construct a state vector that is simultaneously anti-symmetric under all exchanges of fermion pairs, regardless of the spins of other particles not involved in the exchange. The CSP in its standard form is therefore proved false.
\end{enumerate}

\section{Defining Permutation Symmetric State Vectors}
In \cite{York} we showed that the specification of permutation symmetric state vectors hinged on (a) the removal of arbitrary $2\pi$ rotations on spin quantization frames in order to define single-valued state vectors and (b) the selection of separate order independent (symmetrically defined) spin quantization frames for each particle.

\subsection{The Two-Particle Case}
We showed that for a two-particle system, it was impossible to choose a common frame symmetrically between the particles, but a symmetrical choice of separate frames was possible and that these frames were related by a rotation $R_{\mathbf{k}}(\pm\pi)$ where the axis of rotation $\hat{\mathbf{k}}$ is that which bisects the momentum vectors of the two particles:
\begin{equation}\label{eqn:bisect}
\mathbf{k} = \hat{\mathbf{p}}_a + \hat{\mathbf{p}}_b.
\end{equation}

One such choice was the independent helicity frame, which, for the {\em current} particle $c$ is defined by:
\begin{eqnarray}\label{eqn:2parthelframe}
\mathbf{z}_c & = & \hat{\mathbf{p}}_c\nonumber\\
\mathbf{y}_c & = & \hat{\mathbf{p}}_c \times \hat{\mathbf{p}}_o = \hat{\mathbf{p}}_c \times \mathbf{k}
\end{eqnarray}
where $o$ signifies the {\em other} particle.

Another choice we called the "momentum bisecting frame", in which
\begin{eqnarray}\label{eqn:2partbisframe}
\mathbf{z}_c & = & \hat{\mathbf{k}}\nonumber\\
\mathbf{y}_c & = & \hat{\mathbf{p}}_c \times \hat{\mathbf{p}}_o = \hat{\mathbf{p}}_c \times \mathbf{k}
\end{eqnarray}

\subsection{Permutation Symmetric State Vectors For Many Particles}  
In the case of more than two particles, there is, in general, no single axis which bisects the momentum vectors of all particles simultaneously. However, we can generalize eqn. \ref{eqn:bisect} to have the form
\begin{eqnarray}\label{eqn:aggreg}
\mathbf{k} = \sum_i {\hat{\mathbf{p}}_i}
\end{eqnarray}
which reduces to eqn. \ref{eqn:bisect} in the two-particle case. 

We can now generalize the specification of independent helicity frame $H$ for each particle by
\begin{eqnarray}
\mathbf{z}_i & = & \hat{\mathbf{p}}_i\nonumber\\
\mathbf{y}_i & = & \mathbf{k} \times \hat{\mathbf{p}}_i 
\end{eqnarray}
and we generalize the bisecting frame for each particle, to its ``aggregate'' frame:
\begin{eqnarray}
\mathbf{z}_i & = & \mathbf{k}\nonumber\\
\mathbf{y}_i & = & \mathbf{k} \times \hat{\mathbf{p}}_i 
\end{eqnarray}
(In both cases it can also be seen that we have reversed the y-axis compared to the two-particle case of eqns. \ref{eqn:2parthelframe} and \ref{eqn:2partbisframe}, so that the projection of $\mathbf{p}_i$ onto the plane perpendicular to the $z$-axis gives the $x$-axis.)

Using these choices we define the permutation symmetric helicity basis vectors for the symmetrized multi-particle Hilbert space\footnote{See \cite{York} for notation and methodology.}
\begin{eqnarray}
\lefteqn{|Q_a,\mathbf{p}_a,s_a,\lambda_a;Q_b,\mathbf{p}_b,s_b,\lambda_b;Q_c,\mathbf{p}_c,s_c,\lambda_c;...>^H}\nonumber\\
& = & \alpha {\sum\limits_{permutations}}\nonumber\\
& & \hfill |Q_a,\mathbf{p}_a,s_a,\lambda_a>^H |Q_b,\mathbf{p}_b,s_b,\lambda_b>^H |Q_c,\mathbf{p}_c,s_c,\lambda_c>^H ...
\end{eqnarray}
where $\alpha$ is a normalizing scalar, and the permutation symmetric aggregate frame ($A$) basis vectors
\begin{eqnarray}
\lefteqn{|Q_a,\mathbf{p}_a,s_a,m_a;Q_b,\mathbf{p}_b,s_b,m_b;Q_c,\mathbf{p}_c,s_c,m_c;...>^A}\nonumber\\
& = & \alpha {\sum\limits_{permutations}}\nonumber\\
& & \hfill |Q_a,\mathbf{p}_a,s_a,m_a>^A |Q_b,\mathbf{p}_b,s_b,m_b>^A |Q_c,\mathbf{p}_c,s_c,m_c>^A ...
\end{eqnarray}

Note that in neither case, are the spins quantized in the same objective frame. In the helicity case, each spin is quantized in its own helicity frame and in the aggregate frame case, although each spin is quantized along the same {\em axis} $\mathbf{k}$, the orientation of the other axes is separately (and symmetrically) defined for each particle. 

Correspondingly, we can define multi-particle wave functions that are simple products of single-particle wave functions:
\begin{eqnarray}
\lefteqn{\psi^H(Q_a,\mathbf{p}_a,s_a,\lambda_a;Q_b,\mathbf{p}_b,s_b,\lambda_b;Q_c,\mathbf{p}_c,s_c,\lambda_c;...)}\nonumber\\
& = & \chi^H(Q_a,\mathbf{p}_a,s_a,\lambda_a) \chi^H(Q_b,\mathbf{p}_b,s_b,\lambda_b) \chi^H(Q_c,\mathbf{p}_c,s_c,\lambda_c)...
\end{eqnarray}
and
\begin{eqnarray}
\lefteqn{\psi^A(Q_a,\mathbf{p}_a,s_a,m_a;Q_b,\mathbf{p}_b,s_b,m_b;Q_c,\mathbf{p}_c,s_c,m_c;...)}\nonumber\\
& = & \chi^A(Q_a,\mathbf{p}_a,s_a,m_a) \chi^A(Q_b,\mathbf{p}_b,s_b,m_b)
\chi^A(Q_c,\mathbf{p}_c,s_c,m_c)...
\end{eqnarray}

To get to a common canonical frame, we must rotate each particle's spin quantization frame from its independent frame into the common frame. For example
\begin{eqnarray}
\lefteqn{|(Q_a,\mathbf{p}_a,s_a,m_a(R_a))_A;(Q_b,\mathbf{p}_b,s_b,m_b(R_b))_A;...>}\nonumber\\
& = & \alpha {\sum_{permutations}} |Q_a,\mathbf{p}_a,s_a,m_a(R_a)>_A |Q_b,\mathbf{p}_b,s_b,m_b(R_b)>_A ...
\end{eqnarray}
where $R_a$ is the rotation which takes the aggregate frame for $a$ into the canonical frame, and similarly for $b,c...$. If we choose the $z$-axis of our canonical frame to be $\mathbf{k}$, then 
\begin{eqnarray}\label{eqn:canrot}
R_a = R_{\mathbf{k}}(-\phi_a) = R_z(-\phi_a)
\end{eqnarray}
and is the rotation about the canonical $z$-axis which takes the projection of $\mathbf{p}_a$ onto the $x,y$-plane into the $x$-axis. In other words, $\phi_a$ is the azimuthal angle of $\mathbf{p}_a$. As long as $\phi_a,\phi_b,...$ are chosen in a way that doesn't depend on any particular particle ordering, then this state vector will also be permutation symmetric, and related to the aggregate frame state vector by\cite{York}
\begin{eqnarray}\label{eqn:symcan}
\lefteqn{|(Q_a,\mathbf{p}_a,s_a,m_a(R_a))_A;(Q_b,\mathbf{p}_b,s_b,m_b(R_b))_A;...>}\nonumber\\
& = & e^{i(m_a\phi_a + m_b\phi_b + ...)}\ |Q_a,\mathbf{p}_a,s_a,m_a;Q_b,\mathbf{p}_b,s_b,m_b;...>^A
\end{eqnarray}
and the corresponding wave function will be
\begin{eqnarray}
\lefteqn{\psi(Q_a,p_a,\theta_a,\phi_a,s_a,m_a;Q_b,p_b,\theta_b,\phi_b,s_b,m_b;...)}\nonumber\\
& = & \chi(Q_a,p_a,\theta_a,\phi_a,s_a,m_a) \chi(Q_b,p_b,\theta_b,\phi_b,s_b,m_b) ...\nonumber\\
& = & e^{i(m_a\phi_a + m_b\phi_b + ...)}\
\chi^A(Q_a,\mathbf{p}_a,s_a,m_a) \chi^A(Q_b,\mathbf{p}_b,s_b,m_b)...
\end{eqnarray}

We can transform to any other canonical frame, related to C by a rotation, by applying the same rotation to each individual particle separately. Since this introduces no asymmetry between the particles, the resulting state vector will again be exchange/permutation symmetric.

\subsection{Mutual Dependency Amongst Multi-Particle Azimuthal Angles}
One of the consequences of the choice of eqn. \ref{eqn:aggreg} in defining independent quantization frames, is that the rotations of eqn. \ref{eqn:canrot} are not all independent.

To see this, note that eqn. \ref{eqn:aggreg} implies that
\begin{eqnarray}
\sum_i {\bar{\mathbf{p}}_i} = 0
\end{eqnarray}
where $\bar{\mathbf{p}}_i$ is the component of $\hat{\mathbf{p}}_i$ normal to $\mathbf{k}$. Consequently, any of the $\phi$ in eqn. \ref{eqn:canrot} is defined by the others, modulo $2\pi$:
\begin{eqnarray}\label{eqn:tandep}
\tan(\phi_i) = \frac{\sum\limits_{j\ne i} {\sin(\theta_j) \sin(\phi_j)}}{\sum\limits_{j\ne i} {\sin(\theta_j) \cos(\phi_j)}}
\end{eqnarray}

This mutual dependency means that we can define one particle's azimuthal angle in terms of the others and this is what enables us to create order dependent state descriptions. For two particle states, eqn. \ref{eqn:tandep} reduces to the condition that $\phi_b = \phi_a \pm \pi$ (because $\mathbf{k},\mathbf{p}_a,\mathbf{p}_b$ must all lie in the same plane). This interdependence of the two state descriptions is the reason that conventional canonical state vectors are often defined in a way that creates an implicit order dependence if they are to be single-valued. Making such an order dependence explicit enables the derivation of the Pauli Exclusion Principle and its generalization to forbidding states of odd composite spin when all other quantum numbers are identical\cite{Rose}\cite{York} {\em even though it is also perfectly possible to define order independent state vectors for the same physical states}.

\section{Asymmetric State Vectors For Many Particles}  

For any subset $N$ of the particles we can define
\begin{eqnarray}
\mathbf{k}_N = \sum\limits_{i\in N} {\hat{\mathbf{p}}_i}
\end{eqnarray}
and therefore define subset aggregate frames, and a separate common frame $C_N$ for each subset for which $\mathbf{k}_N$ is chosen as the $z$-axis and within each of which 
\begin{eqnarray}
\tan(\phi^{C_N}_i) = \frac{\sum\limits_{(j\ne i)\in N} {\sin(\theta^{C_N}_i) \sin(\phi^{C_N}_j)}}{\sum\limits_{(j\ne i)\in N} {\sin(\theta^{C_N}_i) \cos(\phi^{C_N}_j)}}
\end{eqnarray}

In particular, for any pair of particles $i,j$, we can choose a common frame $C_{ij}$ for which $\phi^{C_{ij}}_j = \phi^{C_{ij}}_i \pm\pi$. In this frame we can define\cite{York} an order dependent state vector that has exchange symmetry $(-1)^{2s_i}$ or $(-1)^{2s_j}$ and, in the limit that all quantum numbers except $m_i,m_j$ are identical, the pair obeys the generalized exclusion rule that states for which the composite (total) spin $S$ is odd are forbidden. This exchange phase and the quantum number $S$ (once we have defined states of definite $S$ in any common frame) are unchanged by rotation to any other common frame. In particular, they are unchanged if we rotate to the multi-particle canonical frame $C$. Hence the exchange asymmetry and the odd $S$ exclusion rule\cite{Rose}\cite{York} will also apply to any pair of particles, for which all other quantum numbers are identical, in this multi-particle canonical frame. 

One could show this process formally, by explicitly relating the multi-particle state vector in canonical frame $C$ to the state vector in which the pair $i,j$ are defined in an order dependent way in the frame $C_{ij}$ to derive a multi-particle canonical state vector which is order dependent for the pair $i,j$. However, as we are about to show, there is a simpler way to get the same result and derive general state vectors that have a variety of exchange asymmetries.

\subsection{Sequence Ranking}\label{sec:orddeprank}
We first of all note that the fact that the dependent $\phi_i$ in eqn. \ref{eqn:tandep} is ambiguous up to integer multiples of $2\pi$, means that eqn. \ref{eqn:tandep} is not sufficient to specify an unambiguous canonical state vector (or wave function) when $s_i$ is half-integer, unless we apply an additional condition such as that it must lie in the range $0 \leq \phi_i < 2\pi$.

We shall define the term {\em rank} ``0'' to describe an azimuthal angle $\phi^0_i$ that has been chosen to satisfy
\begin{eqnarray}\label{eqn:rank_0}
0 \leq \phi^0_i < 2\pi
\end{eqnarray}
Clearly we are free to make the same choice independently for all particles. Hence, from  eqn. \ref{eqn:symcan}, we can define
\begin{eqnarray}
\lefteqn{|(Q_a,\mathbf{p}_a,s_a,m_a)^0;...(Q_i,\mathbf{p}_i,s_i,m_i)^0;...(Q_j,\mathbf{p}_j,s_j,m_j)^0;...>}\nonumber\\
 & = & e^{i(m_a\phi^0_a + ... + m_i\phi^0_i + ... + m_j\phi^0_j + ...)}\nonumber\\
 & & |Q_a,\mathbf{p}_a,s_a,m_a;...Q_i,\mathbf{p}_i,s_i,m_i;...Q_j,\mathbf{p}_j,s_j,m_j;...>^A
\end{eqnarray}
which is symmetric under all exchanges $i\leftrightarrow j$.

We now note that we can also define an azimuthal angle of rank ``1'' by relating it to the azimuthal angle of another particle:
\begin{eqnarray}\label{eqn:rank_1}
\phi^{1,i}_j = \phi^0_i + \Delta_{ji}
\end{eqnarray}
We also choose
\begin{equation}\label{eqn:asymdelta}
\begin{array}{lclcccc}
0\ \leq & \Delta_{ji} & \leq \ 2\pi\\
\Delta_{ij} & = & 2\pi - \Delta_{ji} \\
\Delta_{ji} & = & 0 & \mbox{if} & \phi^0_j = \phi^0_i & \mbox{and} & i<j\\
\Delta_{ji} & = & 2\pi & \mbox{if} & \phi^0_j = \phi^0_i\ & \mbox{and} & i>j
\end{array}
\end{equation}
and define
\begin{equation}\label{eqn:djistrict}
\begin{array}{llllll}
d_{ji} = & 0  & \mbox{if} & \phi^0_j > \phi^0_i\nonumber\\
d_{ji} = & 1  & \mbox{if} & \phi^0_j < \phi^0_i\nonumber\\
d_{ji} = & 0  & \mbox{if} & \phi^0_j = \phi^0_i\ & \mbox{and} & i<j\nonumber\\
d_{ji} = & 1  & \mbox{if} & \phi^0_j = \phi^0_i\ & \mbox{and} & j<i
\end{array}
\end{equation}
from which we see that
\begin{equation}
\begin{array}{rcl}
d_{ij} & = & 1 - d_{ji}\\
\phi^{1,i}_j - \phi^0_j & = & 2\pi d_{ji}\\
\phi^0_j - \phi^0_i & = & \Delta_{ji} - 2\pi d_{ji}\\
\Delta_{ki} & = & \Delta_{ji}+ \Delta_{kj} - 2\pi d_{ji}
\end{array}
\end{equation}

At this point we note that since the particle indeces are essentially arbitrary, we can order them in any way we like. From now on, unless otherwise stated, we shall choose to order our indeces in increasing order of $\phi^0$:
\begin{eqnarray}\label{eqn:indexorder}
\phi^0_j \geq \phi^0_i & \mbox{if} & j>i
\end{eqnarray}
Then we see that eqn. \ref{eqn:djistrict} reduces to the simpler form:
\begin{eqnarray}\label{eqn:dji}
d_{ji} = 0  & \mbox{if} & j>i\nonumber\\
d_{ji} = 1  & \mbox{if} & j<i
\end{eqnarray}

Now, again from  eqn. \ref{eqn:symcan}, if we choose all particles to be of rank 0 except for particle $j$ which we define to be of rank 1 with respect to particle $i$, then we can define
\begin{eqnarray}\label{eqn:01dep}
\lefteqn{|...(Q_i,\mathbf{p}_i,s_i,m_i)^0;...(Q_j,\mathbf{p}_j,s_j,m_j)^{1,i};...>}\nonumber\\
 & = & e^{i m_j (\phi^{1,i}_j - \phi^0_j)}\ |...(Q_i,\mathbf{p}_i,s_i,m_i)^0;...(Q_j,\mathbf{p}_j,s_j,m_j)^0;...>\nonumber\\
 & = & e^{i 2\pi m_j d_{ji}}\ |...(Q_i,\mathbf{p}_i,s_i,m_i)^0;...(Q_j,\mathbf{p}_j,s_j,m_j)^0;...>
\end{eqnarray}
Interchanging $i\leftrightarrow j$,
\begin{eqnarray}
\lefteqn{|...(Q_j,\mathbf{p}_j,s_j,m_j)^0;...(Q_i,\mathbf{p}_i,s_i,m_i)^{1,j};...>}\nonumber\\
& = & e^{i 2\pi m_i d_{ij}}\ |...(Q_i,\mathbf{p}_i,s_i,m_i)^0;...(Q_j,\mathbf{p}_j,s_j,m_j)^0;...>\nonumber\\
& = & e^{i 2\pi (m_i d_{ij} - m_j d_{ji})}\ |...(Q_i,\mathbf{p}_i,s_i,m_i)^0;...(Q_j,\mathbf{p}_j,s_j,m_j)^{1,i};...>
\end{eqnarray}
and, from eqn. \ref{eqn:dji}, we see that the exchange phase reduces to either $(-1)^{2s_j}$ or $(-1)^{2s_i}$, depending on the relative magnitude of $\phi^0_j$ and $\phi^0_i$.

In the limit of equal fermion quantum numbers for both $i$ and $j$, the state vector will therefore vanish. Since this asymmetric state vector is related to the permutation symmetric state vector, or any other state vector for the same state, by at most a phase, all such state vectors must vanish in this limit and we have proved the Pauli principle for any pair of identical fermions in a multi-particle state. Likewise its generalization to the exclusion of odd composite spin $S$ for any pair of identical particles whether fermions or bosons (see \cite{York} for details).

\subsection{Multiple Ranking Sequences}\label{sec:multirank}
We now turn our attention to situations in which more than one particle has its azimuthal angle defined in an order dependent way. We first of all note that if we had allowed any third particle's rank and sequence to depend on either particle $j$ or particle $i$, then we might have obtained a different exchange phase when interchanging $i\leftrightarrow j$ as above, simply because the interchange could have affected the definition of the third particle's azimuthal angle in addition to that of $i$ and $j$.

Extending the ranking system for specifying the azimuthal angles in an order dependent sequence, we can also define an azimuthal angle to be of rank ``2'' when
\begin{equation}
\begin{array}{rclcl}
\phi^{2,ij}_k & = & \phi^{1,i}_j + \Delta_{kj}\nonumber\\
& = & \phi^0_i + \Delta_{ji} + \Delta_{kj} & = & \phi^0_i + \Delta_{ki} + 2\pi d_{ji}\nonumber\\
& = & \phi^{1,i}_k + 2\pi d_{ji} & = & \phi^0_k + 2\pi (d_{ki} + d_{ji})
\end{array}
\end{equation}

In general, the $n$th rank definition of the azimuthal angle with respect to a sequence of $n$ other particles is
\begin{eqnarray}\label{eqn:rank_n}
\lefteqn{\phi^{n,q_1...q_n}_{q_{n+1}} = \phi^{n-1,q_1...q_{n-1}}_{q_n} + \Delta_{q_{n+1}q_n} = \phi^0_{q_1} + \sum\limits_{1<i\leq n} \Delta_{q_{i+1}q_i}}\nonumber\\
& & = \phi^0_{q_1} + \Delta_{q_{n+1}q_1} + 2\pi \sum\limits_{1<i<n} {d_{q_{i+1} q_i}} = \phi^0_{q_{n+1}} + 2\pi N^{q_1...q_{n+1}}
\end{eqnarray}
where $q_i$ is the index of the $i$th particle in the sequence, and
\begin{eqnarray}\label{eqn:rotnum}
N^{q_1...q_{n+1}} & = & d_{q_{n+1}q_1} + \sum_{1<i<n} {d_{q_{i+1} q_i}}\nonumber\\
& = & (1 - 2d_{q_1q_{n+1}}) + (\sum_{1<i<n} {d_{q_{i+1}q_i}} + d_{q_1q_{n+1}})
\end{eqnarray}
The term in the first parentheses is necessarily odd. The term in the second parentheses will be odd(even) if the index sequence $q_{n+1}q_1..q_n$ has an odd(even) number of particle indeces out of index order.

It should be clear from this that whatever ranks and sequences we choose for any particular pair of particles $i,j$, interchanging $i\leftrightarrow j$ will affect the rank and/or the sequence by which their azimuthal angles are defined and similarly for any other particles which have either of these particles in their defining ranking sequence. As a consequence, it will result, in general, in a different rotation by an integer multiple of $2\pi$ on any or all particles. For bosons this will not affect the phase. For fermions it will either result in a sign change or not, depending on how many $2\pi$ rotations are required. To determine whether the sign of the state vector changes, it is necessary to trace back the dependencies for all particles to establish the change in $N^{q_1...q_{n+1}}$. This will be determined by the change in the number of particles which are out of index order due to the change in sequence (if that sequence includes the particles $i$ and/or $j$). The sum of these changes over all fermions will therefore tell us whether or not the state vector is symmetric or anti-symmetric, depending on whether this sum is even or odd. Clearly a wide variety of state vectors can be defined, each of which may or may not change sign under exchange of any particular pair of particles.

One of the questions we must now ask ourselves is: Given this plethora of choices for rank and sequence dependency, is there a set of choices that enables us to define state vectors that have the conventional fermion/boson asymmetry between three or more such particles simultaneously? In other words, can we construct state vectors for more than two identical fermions that obey the CSP?

\section{Emulating The Conventional Symmetrization Postulate}
For convenience, we shall initially consider states of only three particles and we shall choose their index ordering to be $i<j<k$ such that eqn. \ref{eqn:indexorder} is satisfied. We consider the state vector for which each particle is described by rank ``1'' with respect to the previous particle in (cyclic) index order:
\begin{eqnarray}\label{eqn:cyc_3}
|S^k_i;S^i_j;S^j_k> = e^{i2\pi(m_id_{ik} + m_jd_{ji} + m_kd_{kj})}|S_i;S_j;S_k> = (-1)^{2s_i} |S_i;S_j;S_k>
\end{eqnarray}
where we have used the following shorthand:
\begin{eqnarray}
S^k_i & = & (Q_i,\mathbf{p}_i,s_i,m_i)^{1,k}\nonumber\\
S_i & = & (Q_i,\mathbf{p}_i,s_i,m_i)^0
\end{eqnarray}
and the final sign in eqn. \ref{eqn:cyc_3} follows from the definition of eqn. \ref{eqn:dji}. We then see that, interchanging $i\leftrightarrow j$,
\begin{eqnarray}
\lefteqn{|S^k_j;S^j_i;S^i_k> = e^{i2\pi(m_id_{ij} + m_jd_{jk} + m_kd_{ki})}|S_i;S_j;S_k>}\nonumber\\
&& = (-1)^{2s_i + 2s_j} |S_i;S_j;S_k> = (-1)^{2s_j}|S^k_i;S^i_j;S^j_k>
\end{eqnarray}

It is easily seen that exchanging $i\leftrightarrow k$ or $k\leftrightarrow j$ will also all result in exactly one change of sequencing order for the middle particle ($j$) only and therefore in a change of sign if this particle is a fermion. Now, because any single interchange breaks the previous cyclic sequences, we must also examine the consequences of two successive exchanges. It is easy to see that, in fact, a second change will also change the sign by $(-1)^{2s_j}$. For example, $i\leftrightarrow j$ followed by $j\leftrightarrow k$ results in:
\begin{eqnarray}
|S^j_k;S^k_i;S^i_j> = (-1)^{2s_j}|S^k_j;S^j_i;S^i_k> = (-1)^{4s_j}|S^k_i;S^i_j;S^j_k>
\end{eqnarray}
which is, of course, just the original state vector apart from pure permutation. Hence a three fermion state defined in this way is pure anti-symmetric under exchange of any two fermion indeces, since the rotated particle will always be a fermion.

Obviously this is very reminiscent of the CSP. However, there are important differences:
\begin{enumerate}
\item We created a triple cyclic dependency in the state descriptions for the three particles. Exchanging two particle indeces affects the third particle. The CSP in its standard form forbids the presence of a third particle to affect the asymmetry between any pair.
\item The anti-symmetry between two fermions breaks down if the middle particle in $\phi^0$ order is a boson. In that situation, we must resort to the two-particle dependency in eqn. \ref{eqn:01dep}. 
\item This state vector is anti-symmetric under $i\leftrightarrow k$ if both exchanged particles $i,k$ are bosons (even {\em identical} bosons) and the middle particle $j$ in $\phi^0$ order is a fermion! 
\item For a mixture of fermions and bosons, the exchange phase will always depend not so much on the spins of the particles being exchanged, but on the orientation of the canonical frame since this defines which particle will be the ``middle'' particle $j$. (If we rotate the canonical frame about the $z$-axis ($\mathbf{k}$), we can change the $\phi^0$ order, thereby possibly changing the middle indexed particle from a fermion into a boson or {\it vice versa}.)
\end{enumerate}
Note that none of these differences implies any violation of the spin-statistics theorem since the state vector is related by a phase (dependent on the cyclic order in its definition) to the fully permutation symmetric state vector which we have already shown to obey the spin-statistics theorem because of the two-particle asymmetries (when all other particle states are unchanged by the exchange). In particular the third feature arises because the mutual dependencies are such that it is the fermion state that is transformed by the boson exchange (so that even in the limit that the boson states become indistinguishable, the state vectors related by the interchange still differ in their description of the fermion state).

Now consider a multi-particle state in which the individual particle state descriptions have been made with the following dependency rankings:

\begin{enumerate}
\item Index all particles in ascending order of $\phi^0$
\item For all bosons, use rank 0.
\item If there are less than three fermions, use rank 0 for the first and rank 1 with respect to the first for the second, if it is present.
\item If the number of fermions is 3, use rank 1 with respect to the previous fermion in index order for each of them.  
\end{enumerate}

Where the number of fermions is less than two, the symmetry under all interchanges is trivially shown. For two fermions, we have the situation discussed earlier in which there may be a sign change when the rank 1 fermion is exchanged with a boson and there is necessarily a sign change when the two fermions are interchanged. For three or more fermions, we have a situation similar to that just discussed in which there is a sign change for the interchange of any pair of fermions, and a possible sign change if a fermion is interchanged with a boson, but because we are now using rank 0 for all bosons and no fermions have bosons in their ranking sequence, there is no sign change under exchange of two bosons. Hence, as long as we don't exchange any fermions with bosons (which would create a violation of the ranking conditions enumerated above) state vectors defined by these rules, and involving no more than three fermions, will emulate the CSP under all fermion-fermion or boson-boson exchange.

The case of four fermions, however, is slightly more complex. In fact, if we applied our rule 4 to the case of four or more fermions, then under double exchanges the CSP rule would break down, since the second exchange can involve two sequence re-orderings. To see this, consider the index sequence $i<j<k<l$ and define
\begin{eqnarray}\label{eqn:cyc_4}
|..S^l_i..S^i_j..S^j_k..S^k_l..> & = & e^{i2\pi(..m_id_{il}.. + m_jd_{ji}.. + m_kd_{kj}.. + m_ld_{lk}..)}|..S_i..S_j..S_k..S_l..>\nonumber\\
& = & (-1)^{2s_i} |..S_i..S_j..S_k..S_l..>
\end{eqnarray}
We then find that under $i\leftrightarrow j$,
\begin{eqnarray}
|..S^l_j..S^j_i..S^i_k..S^k_l..> & = & (-1)^{2s_j} |..S^k_i..S^i_j..S^j_k..S^k_l..>
\end{eqnarray}
but under the additional exchange $j\leftrightarrow k$, 
\begin{eqnarray}
|..S^l_k..S^k_i..S^i_j..S^j_l..> & = & (-1)^{2s_j + 2s_k} |..S^k_j..S^j_i..S^i_k..S^k_l..>\nonumber\\
& = & (-1)^{2s_k} |..S^l_i..S^i_j..S^j_k..S^k_l..>
\end{eqnarray}
In other words the double exchange is equivalent to a single exchange $k\leftrightarrow l$, which clearly violates the CSP.

However, although we have not been able, using the above rules, to extend the CSP emulation to states of more than three fermions, it is a reasonable conjecture, that it should be possible, using second order rankings to find a state vector involving four fermions that {\em will} emulate the CSP under fermion-fermion or boson-boson exchange. And similarly, using higher rankings when higher numbers of fermions are involved. However, it must be very clear to the reader that such constructions are not only highly complex, but also greatly counter-intuitive and we shall not consider them further in this paper.

\section{Impossibility of the Conventional Symmetrization Postulate In Its Standard Form}
In its standard form, the CSP requires not just simultaneous anti-symmetry under exchange of any pair of identical fermions, but it also requires that the exchange phase for any pair of identical particles, does not depend on the spin of any additional particles that may be present. We shall now prove that it is not possible to define a single state vector that satisfies {\em both} these properties.

For convenience, we shall consider only states of three particles. If we can show that the CSP is false for three fermions, then it must obviously be false for more than three fermions.

We start off, as usual, with a three-particle state vector $|S_i;S_j;S_k> = |S_j;S_i;S_k> = |S_j;S_k;S_i> = ...$ that is simultaneously permutation symmetric for all permutations as long as the individual state descriptions have no order dependence. One example (which we have repeatedly taken advantage of) where we can be sure of this, is when all state descriptions are independent of each other, as in the case of rank 0 in the common canonical frame for which $\mathbf{k}$ is the $z$-axis. We showed in section \ref{sec:orddeprank} that by then creating an order dependence between any pair (e.g. by introducing a higher rank for any particle), we can also define state vectors that are asymmetric for any given pair. 

We saw in section \ref{sec:multirank}, however, that order dependencies can be quite complex. In the three particle case, a single particle's state description could be dependent on either or both other particles. Hence, when interchanging two indeces we can affect not just the state descriptions of the two particles with those indeces but also that of any other particle that depends on some sequence involving either or both of the particles with the exchanged indeces.

To start off simple, we first define a state vector that is separately asymmetric under the exchange of any particular pair:
\begin{eqnarray}\label{eqn:orderedpair}
|S^1_i;S^2_j;S^3_k> = f_{ijk}|S_i;S_j;S_k> = \frac{f_{ijk}}{f_{jik}} |S^1_j;S^2_i;S^3_k>\nonumber\\
|S^2_i;S^3_j;S^1_k> = f_{kij}|S_i;S_j;S_k> = \frac{f_{kij}}{f_{ikj}} |S^2_k;S^3_j;S^1_i>\nonumber\\
|S^3_i;S^1_j;S^2_k> = f_{jki}|S_i;S_j;S_k> = \frac{f_{jki}}{f_{jki}} |S^3_i;S^1_k;S^2_j>
\end{eqnarray}
where the ordering superscripts $1,2,3$ do not necessarily specify a particular rank, but merely that ranks and sequences have been used to create an order dependence in the individual state descriptions and the $f_{ijk}$ are phase factors. Note that eqn. \ref{eqn:orderedpair}, as it stands shows only the results of exchanging particles that have the order labeling ``1'' and ``2''. However, the results of interchanging the pairs ``1'' and ``3'' or ``2'' and ``3'' can also be determined by the phase factors specified in eqn. \ref{eqn:orderedpair}. 

For the CSP in its standard form, however, we must consider only the case where the third particle's state description is unaffected by the order dependence in the other two. If this were not the case, then the exchange could bring about a rotation of the third particle's spin quantization frame, causing the exchange phase to depend on the third particle's spin. To satisfy this requirement we must impose the condition that 
\begin{eqnarray}\label{eqn:indep3}
\frac{f_{ijk}}{f_{jik}} = \eta_{ij} (= \eta_{ji})
\end{eqnarray}

It is apparent then that if we could define the significance of our superscripts $1,2,3$ for three fermions such that $\eta_{ij} = -1$ and, in general,
\begin{eqnarray}\label{eqn:cycasym}
\frac{f_{ijk}}{f_{jik}} = \frac{f_{kij}}{f_{ikj}}= \frac{f_{jki}}{f_{kji}}= -1\nonumber\\
\frac{f_{ijk}}{f_{jki}} = \frac{f_{kij}}{f_{ijk}}= \frac{f_{jki}}{f_{kij}}= +1
\end{eqnarray}
and similarly, giving $-1$($+1$) for the ratios of all odd(even) index permutations, then our state vectors in eqn. \ref{eqn:orderedpair} would obey the CSP.

However, there is no guarantee, in general, that we can define state vectors, even for three fermions, for which the $f_{ijk}$ obey eqn. \ref{eqn:cycasym}, since the $f_{ijk}$ are determined by the physical transformations on the state descriptions that relate the order dependent descriptions $S^1_i$, etc. to the order independent descriptions $S_i$.

As we saw in section \ref{sec:multirank}, these transformations are rotations by multiples of $2\pi$ about the canonical $z$-axis. Given the additional condition (eqn. \ref{eqn:indep3}) of the CSP requirement, in its standard form, that the spin of the unexchanged particle cannot affect the exchange phase, the numbers $N^{q_1...q_{n+1}}$ in eqn. \ref{eqn:rank_n} need to be replaced by numbers of the form $n^p_{q_{n+1}}$ where $p$ indicates the ordering label $1,2,3$ so that, for instance $n^1_i$ is independent of the ordering of $j$ and $k$ and so on. Let us suppose, for the moment, that we can do this. The phase factors $f_{ijk}$ would then take the form:
\begin{eqnarray}\label{eqn:indphase}
f_{ijk} & = & e^{i2\pi(m_i n^1_i + m_j n^2_j + m_k n^3_k)}\nonumber\\
\frac{f_{ijk}}{f_{jik}} & = & e^{i2\pi(m_i (n^1_i - n^2_i) + m_j (n^2_j - n^1_j))}
\end{eqnarray}
To create a three-fermion state vector simultaneously anti-symmetric under all pairwise $i\leftrightarrow j$, $j\leftrightarrow k$ and $k\leftrightarrow i$, then we must simultaneously satisfy (amongst others) at least all three of the following conditions:
\begin{equation}
\begin{array}{rcccl}
\mbox{either} & n^1_i - n^2_i & \mbox{or} & n^2_j - n^1_j & \mbox{must be odd but not both,} \nonumber\\
\mbox{either} & n^1_j - n^2_j & \mbox{or} & n^2_k - n^1_k & \mbox{must be odd but not both and} \nonumber\\
\mbox{either} & n^1_k - n^2_k & \mbox{or} & n^2_i - n^1_i & \mbox{must be odd but not both.} 
\end{array}
\end{equation}
It is easily seen that these three conditions are mutually incompatible. (If $n^1_i - n^2_i$ is odd(even), then to satisfy the first and third conditions, $n^2_j - n^1_j$ and $n^2_k - n^1_k$ must both be even(odd), thereby violating the second condition.) Since it is not possible to satisfy these three conditions simultaneously, then we must conclude that with the pairwise order dependence of eqns. \ref{eqn:cycasym} and \ref{eqn:indphase} (where the third particle is not affected by the interchange, and where the exchange phase for any pair cannot depend on the spins of additional particles which may be present), there can be no single state vector that satisfies eqn. \ref{eqn:orderedpair} simultaneously  for all pairs that can be interchanged.

\section{Summary}
We have shown that it is not possible to define a state vector, for states of three or more identical particles, that obeys the CSP in its standard form. We have also shown that even in its limited form, involving complex multiple dependencies in the state descriptions of the individual particles, the CSP is not a convenient way to summarize the effects of permutation invariance for states of many particles. We therefore conclude that it will almost certainly prove easier to use the purely permutation symmetric state vectors determined by the requirements of the RSP instead, in nearly all circumstances.

\end{document}